\documentclass[pra,twocolumn,superscriptaddress,groupedaddress,showpacs,amsmath,amssymb,floatfix]{revtex4}

\usepackage{epsfig}
\usepackage{graphicx}
\usepackage{dcolumn}
\usepackage{bm}
\usepackage{wasysym}
\def\btt#1{\texttt{\@backslashchar#1}}
\DeclareRobustCommand\bblash{\btt{\@backslashchar}}

\def\>{\rangle}
\def\<{\langle}

\begin{document}

\title{Error-correcting one-way quantum computation with global entangling
gates}

\author{Jaewoo Joo and David L. Feder}
\affiliation{Institute for Quantum Information Science,
University of Calgary, Alberta T2N 1N4, Canada}

\date{\today}

\begin{abstract}
We present an approach to one-way quantum computation (1WQC) that
can compensate for single-qubit errors, by encoding the logical
information residing on physical qubits into five-qubit
error-correcting code states. A logical two-qubit cluster state
that is the fundamental resource for encoded quantum
teleportation is then described by a graph state containing ten
vertices with constant degree seven. Universal 1WQC that
incorporates error correction requires only multiple copies of
this logical two-qubit state and a logical four-qubit linear
cluster state, which are prepared only just in advance of their
use in order to minimize the accumulation of errors. We suggest
how to implement this approach in systems characterized by qubits in
regular two-dimensional lattices for which entangling gates are generically
global operations, such as atoms in optical lattices, quantum dots, or
superconducting qubits.
\end{abstract}


\maketitle

\section{Introduction}
A fundamental requirement for the implementation of reliable
quantum information processing is the ability to diagnose the
presence of errors that might have occurred on quantum bits
(qubits) and to make the appropriate corrections, without
obtaining any knowledge of the quantum information itself.
Quantum error-correcting codes (QECCs) were developed over a
decade ago that can accomplish
this~\cite{Shor95,Calderbank96,Gottesman96,Bennett96,Calderbank97,Knill97},
with a five-qubit QECC being the smallest that can perfectly
protect against an arbitrary error on a single
qubit~\cite{Laflamme96,DiVin96,Knill01}. QECCs serve as a crucial
ingredient for fault tolerance~\cite{Knill05a,Knill05b}, which is
the full protection of quantum information during the
implementation of all quantum gates and measurements, as long as
the frequency of errors is below a certain
threshold~\cite{Knill98,Steane03}. The usual approach to fault
tolerance is through concatenation of QECCs~\cite{Knill96}, though
these generally give very low thresholds~\cite{Aliferis06};
topological approaches to quantum computation can cope in
principle with much higher error
rates~\cite{Kitaev03,Preskill97}. Many QECC schemes have been
successfully demonstrated theoretically and experimentally over
the years~\cite{Benhelm08}.

It remains unclear how best to extend the `one-way' quantum
computation (1WQC) model~\cite{Briegel01,Raussendorf03}, where
processing occurs solely by performing measurements on a highly
entangled `graph' state, to include fault tolerance. The most
intuitive method, which is to implement an encoded quantum
circuit in 1WQC~\cite{Hans,Nielsen05,Aliferis05}, does not
directly correct for errors in the preparation of the relevant
graph state nor errors that accumulate on physical qubits as a
function of time. One approach would be to embed the cluster
state in a decoherence-free
subspace~\cite{Chuang96,Tame2007,Jiang08}. Topological
techniques~\cite{Raussendorf06,Raussendorf07a,Raussendorf07b,Stock08,Devitt08}
solve this problem and yield high error thresholds. A complementary method
for fault-tolerant 1WQC that has been recently proposed is based on embedding
QECC graphs into the computational graph
state~\cite{Silva07,Fujii08,Griffiths}, and our proposal has some
features in common with these. If the encoded graph state is
prepared in advance, however, this approach cannot adequately
compensate for the large number of errors that are likely to have
occurred on distant physical qubits by the time they are measured.

We propose a simple and practical method for 1WQC that incorporates the
five-qubit QECC, which is the minimum size for quantum codes that can correct
single-qubit errors, and which has a simple and intuitive construction in terms
of graph states. The main approach is two-fold. First, rather than
building the full (computational) cluster state in
advance, one instead forms a collection of linear two-qubit and
four-qubit (sideways `horseshoe') cluster states, which together
represent only a small piece of the graph state needed to
simulate a given quantum circuit. In this way, one minimizes the
temporal accumulation of errors on qubits. This strategy also
minimizes the vertex degree of any given vertex in the resulting
graph states, which is advantageous for state purification
protocols~\cite{Dur03,Goyal06}. These features comes at a cost,
however; the entanglement cannot be all generated before the
computation begins as it is in the usual measurement-based model.
This proposal therefore consists of a hybrid of 1WQC and the
quantum circuit method. Second, each of the qubits in these small
cluster states are in fact encoded qubits representing a
five-qubit QECC, so that any single-qubit error that occurs on
any of the (two or four) encoded qubits can be detected and
corrected before any gate teleportation.

Computation proceeds by first encoding the left qubit(s) into the QECC,
performing the encoded entangling gates with the right qubit(s), decoding
the QECC on the left, making the syndrome measurements, and then performing
the desired gate teleportation by measurement. After the gate teleportation
from the left to the right, the logical cluster states are again re-built in
two columns by re-encoding the previously measured qubits on the left, and
entangling these with the logical state encoded on the five (or 10) physical
qubits on the right. The procedure is repeated right-to-left, and then back
again, until the entire computation is accomplished.

In many physical systems, the qubits are naturally arranged in a
regular two-dimensional (2D) lattices, such as ultracold neutral
atoms confined in optical lattices, quantum dots, or charge and
flux superconducting qubits. These systems are often
characterized by local single-qubit measurements, but global
entangling gates, i.e.\ where each qubit becomes entangled with
its nearest neighbors down a given axis. In principle, regular 2D
cluster states, which are a universal resource for
1WQC~\cite{Raussendorf02,Nest06}, can be readily formed
dynamically in principle by applying simple spin Hamiltonians
with quantum
dots~\cite{Borhani05,Weinstein05,Taylor07,Guo07,Lin08},
superconducting qubits~\cite{Tanamoto06,You07,Chen07}, and atoms
in optical lattices~\cite{Duan03,Garcia03,Mandel2003}. Rather
than a complication, global entangling gates can be harnessed in
order to simultaneously encode the qubit(s) on one side while
decoding the qubit(s) on the other. This efficient technique also
allows for a dramatic reduction in the number of entangling
operations required during the whole procedure.

\begin{figure}[t]
\hspace{-1.8cm}
\includegraphics[width= 5.5cm,angle=-90]{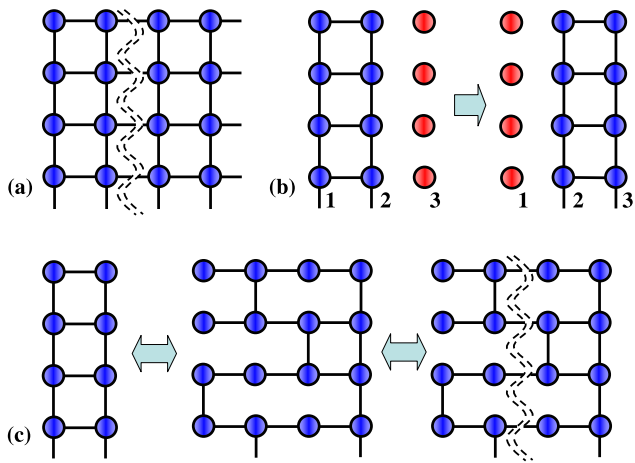}
\vspace{0.8cm} \caption{ \label{New_2DtoLad} (a) A regular 2D cluster
state is comprised of qubits (shown as blue dots) entangled with their nearest
neighbors (solid lines). Alternatively, it can be considered as a two-column
cluster state, in which entangled qubits in column 2 are not yet entangled
with qubits in column 3 (red dots), as shown in (b). Only after measurements
on qubits in column 1 (subsequently shown in red) teleport gates to qubits in
colum 2, the qubits in columns 2 and 3 are entangled, and so on. In (c), the
two-column cluster state is equivalent to a larger 2D cluster state, which
in turn can be split into further two-column cluster states, shown on the
right.}
\end{figure}

A regular 2D cluster state is shown in Fig.~\ref{New_2DtoLad}(a).
Qubits are initialized in the state
$|+\rangle=\left(|0\rangle+|1\rangle\right)/\sqrt{2}$, which is
the positive eigenvector of the Pauli-$X$ operator, and
nearest-neighbor qubits are maximally entangled through the
controlled-phase gate $CZ={\rm diag}\left(1,1,1,-1\right)$. In
the usual approach to 1WQC, measurements are made on qubits
column by column from left to right, which teleport gates in the
same direction. Measurements are made in the Pauli-$XY$ plane
spanned by the vectors $|\pm_{\xi}\rangle=\left(|0\rangle\pm
e^{i\xi}|1\rangle\right)/\sqrt{2}$ corresponding to the operator
$HR_z(\xi)$ where $R_z(\xi)=\begin{pmatrix} e^{-i\xi/2} & 0\cr 0
& e^{i\xi/2}\cr\end{pmatrix}$ generates a rotation about the
Pauli-$Z$ axis. On outcome $m\in\{0,1\}$, the measurement has the
effect of teleporting the gate $X^mHR_z(\xi)$, where $H$ is the
Hadamard operator. Three of these gates with different choices of
$\xi$ are sufficient to simulate a general single-qubit unitary.
Together with the vertical $CZ$ links that simulate entangling
gates between logical qubits, 1WQC is able to simulate any
quantum circuit, and is one of many approaches in the general
measurement-based model of quantum computation~\cite{Aliferis04}.

Because of the linearity of the $CZ$ gates, the computation on
the full 2D cluster can be instead constructed as a sequence of
teleportations and column-to-column entangling operations, as
shown in Fig.~\ref{New_2DtoLad}(b). Qubits in the first two
columns are first entangled, and a gate teleportation is effected
by measurements of qubits in column 1. The qubits in the second
column are only then entangled with those in column 3, and the
procedure is repeated. Evidently, only two columns of qubits are
ever actually needed in order to reproduce the full 2D cluster.
The qubits in the left column are initially measured and the gate
is teleported to qubits on the right; after re-entangling the two
columns, measurement of qubits on the right will teleport another
gate back to qubits on the left, etc. A similar idea has been recently
discussed in Ref.~\cite{Raussendorf07a,Wunderlich}.

As illustrated in Fig.~\ref{New_2DtoLad}(c), the 2D cluster state
can always be decomposed into a series of two-column graph states
that are characterized by only two different kinds of subgraphs.
These correspond to a two-qubit graph state $|g_-\rangle\equiv
CZ_{1,2}|++\rangle_{1,2}$ oriented horizontally, and the
four-qubit linear cluster state in the `horseshoe' shapes
$\sqsubset$ or $\sqsupset$,
$|g_{\sqsubset}\rangle=|g_{\sqsupset}\rangle\equiv
CZ_{1,2}CZ_{2,3}CZ_{3,4}|++++\rangle_{1,2,3,4}$. This
decomposition is always possible because inserting additional
qubits along a horizontal axis simply requires additional
measurements in the $X$ basis, which teleport trivial (Clifford
group) operators. Evidently, the same decomposition is also
possible with computational cluster states, in which physical
qubits are removed from the 2D cluster by computational basis
(Pauli-$Z$) measurements. Thus, to effect universal quantum
computation, one requires only multiple copies of the two states
$|g_-\rangle$ and $|g_{\sqsubset}\rangle$.

The remaining ingredient, and the core of the present work, is to encode each
physical qubit in the states $|g_-\rangle$ and $|g_{\sqsubset}\rangle$ into
QECC graph states. To minimize the overhead in terms of physical qubits, and
the complexity of forming the states, we focus on the five-qubit QECC.
A logical qubit $A$ with five physical qubits (labeled 1 through 5)
is represented by~\cite{Bennett96}
\begin{eqnarray}
\label{eq:LQ01} | 0^{L} \rangle_{A} &=& {1\over4} \big( |00000
\rangle + |10010 \rangle + |01001 \rangle + |10100 \rangle \nonumber \\
&& \,+ |01010 \rangle - |11011 \rangle - |00110 \rangle - |11000 \rangle \nonumber \\
&& \,- |11101 \rangle - |00011 \rangle - |11110 \rangle - |01111 \rangle \nonumber \\
&& \,- |10001 \rangle - |01100 \rangle - |10111 \rangle + |00101
\rangle \big)\,,~~~~~
\end{eqnarray}
and the other logical qubit on site $A$ is equal to
\begin{eqnarray}
\label{eq:LQ02} | 1^{L} \rangle_{A} &=& X^{L}_{A} | 0^{L}
\rangle_{A} = X^{\otimes 5}_{1-5}| 0^{L}\rangle_{A},~~~
\end{eqnarray}
where $X^{L}_{A}$ and $Z^{L}_{A}=Z^{\otimes 5}_{1-5}$ are logical Pauli-$X$
and $Z$ operations on the logical qubit $A$ and are fully transversal (see
details in Ref.~\cite{Nielsen}).
Then, the eigenvectors of $X^L_A$, hitherto referred to as logical Hadamard
states, are also defined by
\begin{eqnarray}
\label{eq:LQ03} | \pm^{L} \rangle_{A} &=& ( | 0^{L} \rangle_{A}
\pm | 1^{L} \rangle_{A} ) / \sqrt{2}\,.~~~
\end{eqnarray}
These are equivalent to a pentagon graph state:
\begin{eqnarray}
| -^{L} \rangle_{A}&\equiv&|\pentagon\rangle_A = C^{\pentagon}_{1-5} | +
\rangle^{\otimes 5}_{1-5} ,\label{eq:minusL} \\
| +^{L} \rangle_{A}&\equiv&|\tilde{\pentagon}\rangle_A=Z^L|\pentagon\rangle_A
=C^{\pentagon}_{1-5}| -\rangle^{\otimes 5}_{1-5},\label{eq:plusL}
\end{eqnarray}
where $|\pm\rangle^{\otimes 5}_{1-5} = |\pm\rangle_{1}|\pm\rangle_{2}
|\pm\rangle_{3}|\pm\rangle_{4}|\pm\rangle_{5}$ and
$C^{\pentagon}_{1-5} = CZ_{1,2}\, CZ_{2,3}\, CZ_{3,4}\, CZ
_{4,5}\, CZ_{5,1}$.
If needed, the logical computational basis states can also be obtained
directly from these via $|1^L\rangle_{A}=X^L_{A}|0^L\rangle_{A}$ and
\begin{eqnarray}
|0^L\rangle_{A} &=& H_1X^{\otimes
4}_{2-5}\prod_{m=3}^4CZ_{1,m}CZ_{2,m}CZ_{5,m}
|-^L\rangle_A\nonumber \\
&=& H_1X^{\otimes 4}_{2-5}CZ_{2,3}CZ_{2,5}CZ_{4,5}|K_5\rangle_{A},
\end{eqnarray}
neglecting an overall sign, where
$|K_5\rangle_{A}=\prod_{m=1}^4\prod_{n=m+1}^5CZ_{m,n}
|+\rangle^{\otimes 5}_{1-5}$ is the complete graph on five qubits.

\begin{figure}[t]
\hspace{-1.8cm}
\includegraphics[width= 5.5cm,angle=-90]{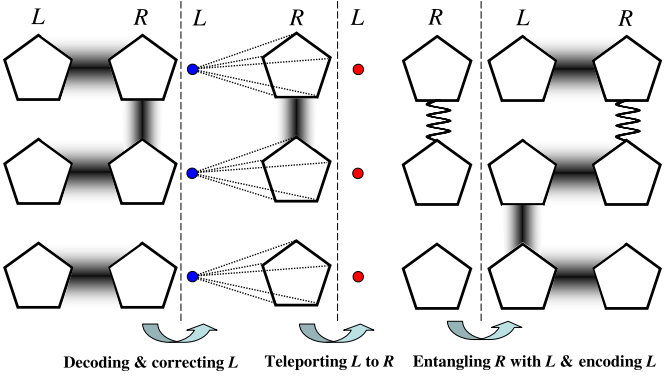}
\vspace{-0.5cm} \caption{ \label{EC1WQC01} A scheme for 1WQC is
depicted in a two-column cluster state (pentagon; logical qubit
and blurred line; logical $CZ$ operation). One performs decoding on
logical qubits and check error syndromes by measurement on
ancillary physical qubits in column $L$. The error-corrected
qubits are teleported to logical qubits in column $R$ (red dot;
measured physical qubit and wiggled line; preexisted $CZ$
operation). Logical qubits are re-built in column $L$ as long as
logical $CZ$ operations are performed among the logical qubits, and
then it is ready to repeat all the procedure from the right to the
left.}
\end{figure}

Fig.~\ref{EC1WQC01} shows a schematic for 1WQC in the two-column
format ($L$ and $R$) discussed above, for a decomposed 2D cluster
state. This particular example corresponds to the first two
columns and first three rows of the third graphic shown in
Fig.~\ref{New_2DtoLad}(c). In the first step, logical cluster
states are prepared in two columns. The pentagon shape denotes a
logical qubit in the state $|\pm^L\rangle$ (the choice of
$|\pentagon\rangle$ or $\tilde{\pentagon}\rangle$ is arbitrary
because the $Z^L$ and $CZ$ gates commute). A blurred line
represents a logical $CZ$ operation (defined formally in the next
Section). After decoding the logical qubits and performing
syndrome measurements on four of the five qubits comprising each
$|\pentagon\rangle$ on the left (second step), a teleportation
scheme performed by single-qubit measurements on the fifth qubits
in column $L$ transfers the quantum information to the logical
qubits in column $R$ (third step in Fig.~\ref{EC1WQC01}). Last,
logical $CZ$ operations are carried out to re-encode the physical
qubits in column $L$, and to perform the various logical $CZ$
operations according to the computational cluster being
simulated. The full procedure is then repeated from right to
left, and then back again, until the desired computation is done.

The proposed scheme is thus a hybrid of the 1WQC and quantum circuit models,
combining the main advantages of both while minimizing the disadvantages. The
main apparent advantage in the 1WQC model, that all of the entanglement can
be generated in advance, is not in fact applicable when the probability of
single-qubit errors is assumed to be constant with time: distant portions of
the cluster state will be heavily distorted by the time measurements are made.
By generating entanglement in only two columns at a time, this problem is
mitigated. Meanwhile, the main advantage of standard 1WQC,
that the nearest-neighbor entangling can be performed using a single global
operation, is preserved in the current approach. Yet effective arbitrary
two-qubit gates can still be performed, in the spirit of the circuit model.
That said, some parallelism of the 1WQC model caused by the sequential
error-checking stage is inevitable in the encoding and decoding procedures,
but the total number of required operations is much smaller than it would be
in the standard circuit model, as discussed in more detail below.

The remainder of this paper is organized as follows. In
Sec.~\ref{QECC_theory}, the mathematical formalism for the
formation of QECC cluster states is described, and the procedure
for implementing 1WQC is discussed. A practical approach to
implementing these ideas for systems characterized by local
single-qubit gates but global entangling operations is shown in
Sec.~\ref{implementation}, and the results are analyzed in
Sec.~\ref{conclusions}.

\section{Theory of 1WQC using 5-qubit QECC}
\label{QECC_theory} The full theoretical approach for 1WQC with
embedded five-qubit quantum error correction can now be
presented. The $CZ$ operation can be represented as
\begin{eqnarray}
\label{eq:CZ01} CZ_{i,j} = {1\over2} \left( {\rm I}_{i} {\rm
I}_{j} + {\rm I}_{i}  Z_{j} + Z_{i}  {\rm I}_{j} - Z_{i} Z_{j}
\right),
\end{eqnarray}
so that a physical two-qubit cluster state between qubits 1 and 2 is
equivalent to
\begin{eqnarray}
\label{eq:CZ02} CZ_{1,2} \, |+\rangle_{1} |+\rangle_{2} =
{1\over\sqrt{2}} ( | 0 \rangle_{1}| + \rangle_{2} + | 1
\rangle_{1}| - \rangle_{2}).
\end{eqnarray}
Likewise, the logical $CZ$ gate can be represented as
\begin{eqnarray}
\label{eq:LCZ01} CZ^{L}_{A,B} = {1\over 2} \left({\rm I}^{L}_{A}
{\rm I}^{L}_{B} + {\rm I}^{L}_{A} Z^{L}_{B} + Z^{L}_{A} {\rm
I}^{L}_{B} - Z^{L}_{A} Z^{L}_{B}\right),
\end{eqnarray}
where ${\rm I}^{L}_{A} = {\rm I}^{\otimes 5}_{1-5}$, so that the logical
two-qubit cluster state is
\begin{equation}
\label{eq:LCZ02} CZ^{L}_{A,B} | -^{L} \rangle_{A}| -^{L}
\rangle_{B} = {1 \over \sqrt{2}} \left( | 0^{L} \rangle_{A}| -^{L}
\rangle_{B} - | 1^{L} \rangle_{A}| +^{L} \rangle_{B}\right).
\end{equation}

\begin{figure}[t]
\hspace{-1.8cm}
\includegraphics[width= 6.3 cm,angle=-90]{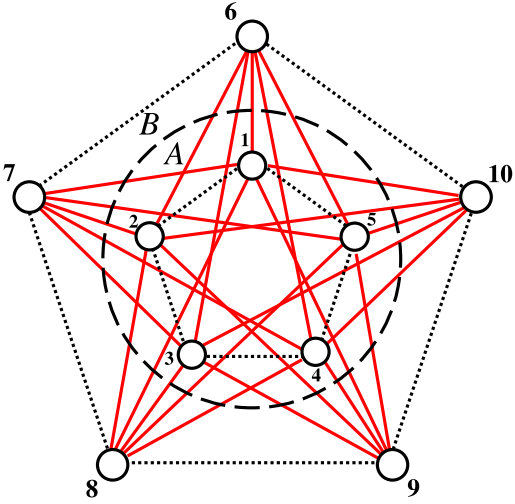}
\vspace{1cm} \caption{ \label{10qubit01} A logical two-qubit
cluster state $|LCS_{2}\rangle_{AB}$ is depicted (see
Eq.~\ref{eq:LCZ03}). Inner (outer)-pentagon qubits correspond to
logical qubit $A$ ($B$) while the red lines indicate physical $CZ$
operations between physical qubits.}
\end{figure}

Of course, the logical $CZ$ gate~(\ref{eq:LCZ01}) is impossible to
implement directly. Rather, one needs to know how to obtain the
resulting state~(\ref{eq:LCZ02}) using only two-qubit $CZ$ gates
and possibly local unitaries. To make further progress, it is
useful to note that a six-qubit GHZ state can be constructed by
adding a sixth qubit initialized in the $|+\rangle$ state as
follows:
\begin{eqnarray}
\label{eq:LQ05} |6GHZ\rangle_{1-5,6} &=& \left[ \prod^{5}_{n=1}
CZ_{n,6}\right] | + \rangle^{\otimes 5}_{1-5}| + \rangle_{6} \\
&=& { 1 \over \sqrt{2}} \left( | + \rangle^{\otimes 5}_{1-5} |
0 \rangle_{6}  + | - \rangle^{\otimes 5}_{1-5} | 1 \rangle_{6}
\right).\nonumber
\end{eqnarray}
Applying the operation $C^{\pentagon}_{1-5}$ on the GHZ state, a
logical-physical cluster state $|\psi_{LP}\rangle$ is formed
\begin{eqnarray}
|\psi_{LP}\rangle &\equiv& C^{\pentagon}_{1-5}|6GHZ\rangle_{A,6}
\nonumber \\
&=& {1\over \sqrt{2}}\left[ | -^{L} \rangle_{A} | 0 \rangle_{6}  +  | +^{L}
\rangle_{A} | 1 \rangle_{6}) \right],
\label{eq:LPCS}
\end{eqnarray}
which is equivalent to the state (\ref{eq:CZ02}) with the substitution
$|+\rangle\to|-^L\rangle$ as shown in Eq.~(\ref{eq:minusL}). By a
straightforward extension of the above arguments, one can readily obtain a
logical two-qubit cluster state with 10 physical qubits:
\begin{eqnarray}
\label{eq:LCZ03} |LCS_2\rangle &\equiv & C^{\pentagon}_{1-5}
C^{\pentagon}_{6-10}\prod^{5}_{m=1}\prod^{10}_{n=6}CZ_{m,n}
|+\rangle^{\otimes 5}_{1-5}|+\rangle^{\otimes 5}_{6-10}\nonumber \\
& = & {1\over \sqrt{2}}\left[ | -^{L} \rangle_{A}
| 0^L \rangle_{B}  +  | +^{L} \rangle_{A} | 1^L \rangle_{B}) \right].
\end{eqnarray}
That this is equivalent to (\ref{eq:LCZ02}) can be seen by noting that
$\prod^{5}_{m=1}\prod^{10}_{n=6}CZ_{m,n}|+\rangle^{\otimes 5}_{1-5}
|+\rangle^{\otimes 5}_{6-10}$ yields
\begin{equation}
{1\over 2} \left[ (| + \rangle^{\otimes 5}_{A}+|
- \rangle^{\otimes 5}_{A}) | + \rangle^{\otimes 5}_{B} + (| +
\rangle^{\otimes 5}_{A}-| - \rangle^{\otimes 5}_{A} ) | -
\rangle^{\otimes 5}_{B}\right],\nonumber
\end{equation}
which due to relations (\ref{eq:minusL}) and (\ref{eq:plusL}) transforms into
(\ref{eq:LCZ02}) upon the application of the $C^{\pentagon}_{1-5}
C^{\pentagon}_{6-10}$ operators.

%and similar mathematical approaches have been investigated for
%different interests in Ref. \cite{Griffiths}.

The encoded two-qubit cluster state is depicted in
Fig.~\ref{10qubit01}. To build this state, all physical
operations are decomposed by two groups of $CZ$ operations. Two
logical qubits $A$ (1 to 5) and $B$ (6 to 10) are linked with
five GHZ-type connections and two pentagon operations, and these
$CZ$ operations importantly commute with each other. The resulting
ten-vertex graph has constant vertex degree seven. Likewise, the
encoded version of the horseshoe subgraphs $\sqsubset$ and
$\sqsupset$ are twenty-vertex graphs, with ten of the vertices
(correspoding to encoded vertices at the two endpoints of the
linear cluster state $g_{\sqsubset}^L$) having degree seven, and
the remaining vertices (encoded vertices in the interior of the
cluster state) having degree twelve.

\begin{figure}[t]
\hspace{-1.8cm}
\includegraphics[width= 5.7cm,angle=-90]{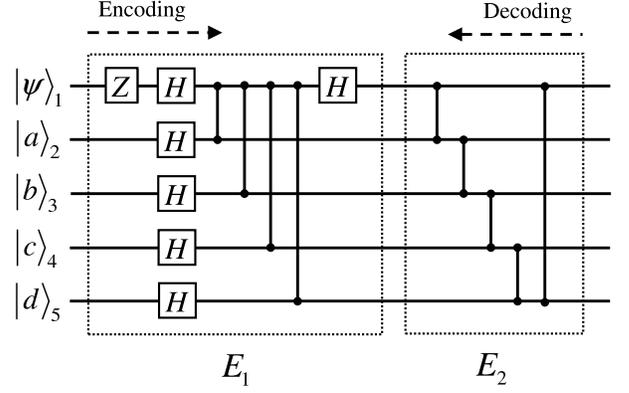}
\caption{ \label{Encoder} For encoding $|\psi\rangle_1 =
\alpha|0\rangle_1 + \beta |1\rangle_1$ into a logical qubit, qubits $2-5$
are prepared in $|0000\rangle_{2345}$. After operations $E_1$ and $E_2$, the
final state is $\alpha |0^{L} \rangle_1 + \beta |1^{L}\rangle_1$.}
\end{figure}

\subsection{Encoding and decoding circuits}
\label{theory3}

While $C^{\pentagon}$ is sufficient to effect the transformation
$|+\rangle^{\otimes 5}\to|-^L\rangle$, after one gate teleportation the
resulting state will instead be $|\psi^L\rangle=\alpha|0^L\rangle
+\beta|1^L\rangle$, with $\alpha,\beta$ arbitrary complex coefficients.
Furthermore, under certain circumstances it may be desirable to initialize
the 1WQC with some particular quantum state. The quantum circuit for encoding
a physical qubit in the state $|\psi\rangle_1=\alpha|0\rangle+\beta|1\rangle$
into a logical qubit consisting of five physical qubits is shown in
Fig.~\ref{Encoder}. The encoding consists of two sets of operations, labelled
$E_1$ and $E_2$. Suppose that five physical qubits are initially prepared in
the state $|\psi\rangle_1 |0000\rangle_{2-5}$. After $E_1$, one obtains
the five-qubit state
\begin{eqnarray}
\label{eq:LQ06} |\tilde{\psi}\rangle_{1-5} = {1\over \sqrt{2}} \left[
(\alpha-\beta) | + \rangle^{\otimes 5}_{1-5} +(\alpha+\beta) |
- \rangle^{\otimes 5}_{1-5}\right] .~~
\end{eqnarray}
The second circuit $E_2$ maps vectors $|\pm\rangle^{\otimes 5}$ into logical
Hadamard states, as discussed in the previous section. Thus, the final state
becomes
\begin{eqnarray}
\label{eq:LQ07}  |\psi^L\rangle_{1-5}=C^{\pentagon}_{1-5}
|\tilde{\psi}\rangle_{1-5} = \alpha | 0^{L} \rangle_{1-5}
+ \beta | 1^{L} \rangle_{1-5}.
\end{eqnarray}
Evidently, for the special case $|\psi\rangle_1=|+\rangle_1$ where
$\alpha=\beta=1/\sqrt{2}$, one obtains
$E_2E_1|+\rangle^{\otimes 5}_{1-5}=|+^L\rangle_{1-5}=Z^LC^{\pentagon}
|+\rangle^{\otimes 5}_{1-5}$.

\begin{table}[b]
\centering
\begin{tabular}{ccc} \hline
~~~Error type~~~ & ~~~Syndrome ($|a\,b\,c\,d\rangle_{2345}$)~~~ & Outcome~~~ \\
\hline
None & 0000 &  \\
$Z_{2}$ & 1000 & \\
$Z_{3}$ & 0100 & $|\psi \rangle$ \\
$Z_{4}$ & 0010 & \\
$Z_{5}$ & 0001 &  \\\hline
$X_1$ & 1001 &  \\
$X_3$ & 1010 &  \\
$X_4$ & 0101 & $X\,|\psi \rangle$ \\
$X_{3}Z_{3}$ & 1110 &  \\
$X_{4}Z_{4}$ & 0111 &  \\ \hline
$X_{1}Z_{1}$ & 0110 &  \\
$X_2$ & 1011 & \\
$X_5$ & 1101 & $XZ\,|\psi \rangle$ \\
$X_{2}Z_{2}$ & 0011 & \\
$X_{5}Z_{5}$ & 1100 & \\ \hline $Z_{1}$ & 1111 & $Z\,|\psi
\rangle$ \\ \hline
\end{tabular}
\caption{Error correction table with corresponding syndrome
detections in the decoder circuit in Fig.~\ref{Encoder}.
}\label{Table1}
\end{table}

The decoding is simply the adjoint of the encoding circuit, as usual. For
example, the detection outcome in $|abcd\rangle_{2345}$ becomes
$|0000\rangle_{2345}$ without an error, because decoding is an
inverse process of encoding qubits. Otherwise, the detected
errors in $|abcd\rangle_{2345}$ represent the physical qubit that suffered
a Pauli error, and which type of error occurred (see the details in
Table~\ref{Table1}). Thus, quantum error-correction is perfectly possible with
corresponding syndrome measurements.

\subsection{Encoded teleportation}
\begin{figure}[t]
\hspace{-1.8cm}
\includegraphics[width= 5.7cm,angle=-90]{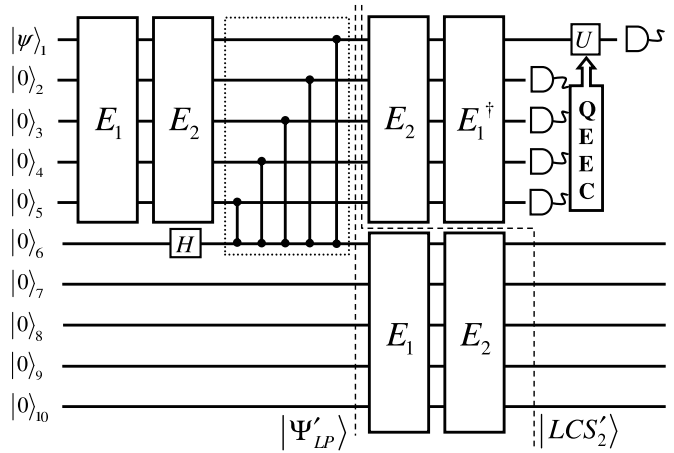}
\vspace{1cm} \caption{The full quantum circuit is shown for encoded
teleportation between the first encoded qubit (physical qubits $1-5$) and
the second (qubits $6-10$). Encoding is accomplished by the unitaries $E_1$
and $E_2$ shown in Fig.~\ref{Encoder}. The two logical encoded qubits are then
entangled by a GHZ-type interaction between the first five physical qubits and
qubit 6. The first logical qubit is decoded via $E_2^{\dag}$ and $E_1^{\dag}$,
syndrome measurements on qubits $2-5$ are performed, and teleportation is
finally accomplished via measurement of qubit 1.}
\label{Teleport}
\end{figure}

An encoded (or error-correcting) teleportation with two logical
qubits is one of the main building blocks of encoded 1WQC. In
Fig.~\ref{Teleport}, the first qubit $|\psi\rangle_1$ has quantum
information and nine physical qubits are initially prepared in
the state $|0 \rangle$. The encoding operations ($E_1$ and $E_2$)
on the first five qubits then yield logical qubit $A$.
Alternatively, the first five qubits would represent a logical
state $|\psi^L\rangle_{1-5}$ that was the result of some previous
, in which case one would ignore the $E_2E_1$ operation. In the
middle of Fig.~\ref{Teleport}, a GHZ operation $\prod^{5}_{n=1}
CZ_{n,6}$ including a Hadamard operation in qubit 6 creates the
logical-physical cluster state
\begin{equation}
\label{LP01} |\psi'_{LP}\rangle =
\alpha|0^L\rangle_{1-5}|+\rangle_6 +
\beta|1^L\rangle_{1-5}|-\rangle_6.
\end{equation}
The subsequent encoding of logical qubit $B$ yields the desired final state
\begin{eqnarray}
\label{eq:Teleport01}  |LCS'_{2}\rangle_{AB} = \alpha | 0^{L}
\rangle_A | +^{L} \rangle_B + \beta | 1^{L} \rangle_A | -^{L}
\rangle_B\,.
\end{eqnarray}

Note that the quantum circuit shown in Ref.~\ref{Teleport} requires only the
single GHZ operation $\prod^{5}_{n=1} CZ_{n,6}$. This is in apparent contrast
with the logical two-qubit state depicted in Fig.~\ref{10qubit01}. In fact,
if one were to `push' the $E_1$ and $E_2$ operations on qubits
6-10 through this GHZ operation, one would indeed recover the implied
graph-state connections. More precisely, it is straightforward to verify that
\begin{eqnarray}
&&E_2E_1\prod^{5}_{n=1} CZ_{n,6}H_6|0\rangle^{\otimes 5}_{6-10}\nonumber \\
&&\qquad =Z^L_{6-10}C^{\pentagon}_{6-10}\prod^{5}_{n=1}\prod_{m=6}^{10}CZ_{n,m}
|+\rangle^{\otimes 5}_{6-10}.\label{eq:push}
\end{eqnarray}
It is clearly preferable in practice to implement the quantum
circuit shown in Fig.~\ref{Teleport} than to explicitly perform
all of the GHZ operations on the right-hand side of
Eq.~(\ref{eq:push}).

Now, quantum information is protected from a single-Pauli error in
the logical two-qubit cluster state. After decoding and measuring
syndromes in logical qubit $A$, the error can be corrected in
qubit 1, yielding the intermediate physical-logical state
\begin{eqnarray}
\label{eq:Teleport02}  |\psi_{PL}\rangle_{1,B} = \alpha | 0 \rangle_1
| +^{L} \rangle_B + \beta | 1 \rangle_1 | -^{L} \rangle_B.
\end{eqnarray}
It is easy to check that a measurement of the first qubit in the $HR_z(\xi)$
basis with outcome $m\in\{0,1\}$ yields the result
\begin{eqnarray}
|\psi^L_{\rm out}\rangle&=&\frac{e^{-i\xi/2}}{\sqrt{2}}\left[\left(
\alpha\pm e^{i\xi}\beta\right)|0^L\rangle_B+\left(\alpha\mp e^{i\xi}\beta\right)
|1^L\rangle_B\right]\nonumber \\
&=&\left(X^L\right)^mH^LR_z^L(\xi)|\psi^L\rangle_B,
\end{eqnarray}
which yields precisely the desired gate on the encoded qubit $B$ required to
construct a universal unitary operation on the logical state. The
generalization of the above results to the encoded horseshoe subgraph is
straightforward.

\section{Physical implementation}
\label{implementation}

Even with the simplified quantum circuit for encoding the
two-qubit cluster state shown in Fig.~\ref{Teleport}, compared
with the full graph state shown in Fig.~\ref{10qubit01}, there
remain 23 physical two-qubit gates in order to implement the
logical two-qubit cluster state. The situation with the encoded
linear four-qubit states is worse yet, with 51 physical two-qubit
gates. The number of such operations can be greatly reduced,
however, in physical systems characterized by global entanglement
operations, such as quantum
dots~\cite{Borhani05,Weinstein05,Taylor07,Guo07,Lin08},
superconducting qubits~\cite{Tanamoto06,You07,Chen07}, and atoms
in optical lattices~\cite{Duan03,Garcia03,Mandel2003}.

Several proposals for the generation of cluster states
using global interactions have been theoretically studied in
solid-state systems. A single quantum-dot qubit consists of two charge 
states in a double-well potential coupled to a long transmission 
resonator~\cite{Lin08}. Two
electrons can be located in either the left or right potential
well. When the double-well is biased by an external field, the charge 
states can be encoded as computational qubit states. The application of an
oscillating field to each qubit then yields a linear cluster 
state~\cite{Lin08,Guo07}, or an encoded cluster state with multiple 
dots~\cite{Weinstein05}.  Similarly, a superconducting qubit consists of 
Josephson junctions connected to a common inductance. For
example, a superconductor ring consisting of three Josephson
junctions can provide a phase qubit in a double-well potential in
which circulating supercurrents of opposite circulation are
computational states. By turning on the inter-qubit inductive
coupling, a linear cluster state can be created in the
superconductor array~\cite{Tanamoto06,You07,Chen07}.

As the other concrete example discussed in the remainder of this work,
consider ultracold atoms confined in 2D optical lattices. Ideally in the
Mott-insulator state~\cite{Jaksch99,Greiner02,Zhang07}, exactly one atom will
occupy each site of a 2D lattice. The Mott limit has been
achieved for arbitrary dimensions~\cite{Kohl05}, though in the 1D
and 2D cases it is difficult to reach the regime of unit
filling~\cite{Stock05,Spielman07,Spielman08}. Entanglement
between each nearest-neighbor can then be effected across the
system using either state-dependent collisions~\cite{Jaksch99,Mandel2003} 
or tunable spin-spin interactions~\cite{Duan03}. In both cases, the 
entanglement operation is global, i.e.\ is effected between all neighbors
simultaneously along a given direction. 
Errors in the application of the global entangling gate will
generally result in a two-qubit controlled-phase gate ${\rm
diag}(1,1,1,e^{i\phi})$ with $\phi\neq\pi$, which would have
serious consequences for the application of 1WQC in these
systems. A global phase error can be eliminated in
principle~\cite{Garrett2008}, but random phase errors are more
difficult to correct~\cite{Tame2005}. 

Only using state-dependent collisions, however, can the choice of 
neighbors be easily
controlled by suitable time-dependent manipulation of the lattice
potential in both spatial directions. In order to apply this
idea to our QECC scheme, every five qubits should be encoded as a
logical qubit in optical lattices. In particular, construction
of all encoded two- or four-qubit cluster states in a given
column of the full computational cluster requires at most five
entangling operations. Furthermore, the decoding circuit
$E_1^{\dag}E_2^{\vphantom{\dag}}$ on the first logical qubit,
shown in Fig.~\ref{Teleport}, can be performed simultaneously
with the encoding circuit $E_2E_1$ on the second qubit. Thus,
each gate teleportation for all logical qubits requires a total
of five entangling gates.

The actual implementation of 1WQC still remains experimentally challenging
because the distance between adjacent lattice sites is usually
comparable to the spatial size of the laser beam used for
single-qubit operations. This means that applying single-qubit
rotation might yield undesirable operations on neighboring atoms
of the target atom using a single laser field. Several proposals
for improving the addressability have recently been made. For
example, interference of laser beams allows single-qubit
operations with many-atom addressing~\cite{manyaddress}; other
proposals include employing microwave transitions~\cite{Sarma} or
pointer atoms~\cite{pointer}. Much experimental progress has been made 
toward single-atom addressing~\cite{Bloch08}. One approach is to 
make a larger separation between the neighboring atoms.
A superlattice scheme using lasers at two different wavelengths enables 
loading of atoms at every third site of the optical 
lattice~\cite{Porto03}. More recently, images of single atoms in 3D optical 
lattices can be taken either through the use of a high-resolution 
lens~\cite{Grangier07} or lattices with large spacing~\cite{Weiss07}. 
Alternatively, atoms loaded into optical lattices~\cite{Nelson07} can be 
rearranged by using two crossed beams~\cite{Miroshnychenko06} to end up
in regions more amenable to addressing.

In this section, we show how to generate logical two-qubit and
four-qubit cluster states in such a system. The main idea is to
use auxiliary physical qubits (ancillae) in the vicinity of the
qubits comprising the logical states, in order to effect
entanglement between two spatially distant physical qubits. We
also illustrate that encoding and decoding can be performed
simultaneously on different logical qubits using global $CZ$
operations, allowing for a significant reduction in the number of
overall operations. Finally, we suggest various strategies for
approaching fault-tolerance in these systems.

\begin{figure}[t]
\hspace{-1.8cm}
\includegraphics[width= 6cm,angle=-90]{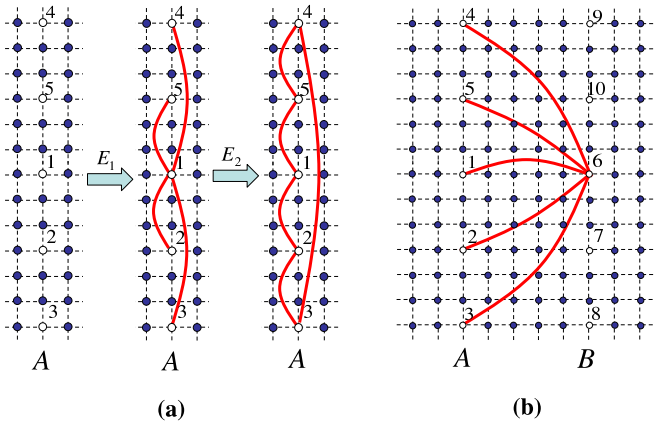}
\caption{ \label{E1E2GHZ} An overview of the procedure for producing the
logical-physical cluster state in a regular lattice is shown. The encoding
operation on the five physical qubits (white dots aligned in a single column)
of logical qubit $A$ is shown in (a). Blue dots denote
auxiliary qubits which mediate a $CZ$ operation (red line) between two distant
qubits. Pauli-$Z$ and Hadamard operations are first performed on the
information qubit $|\psi\rangle_1$ (see Fig.~\ref{Encoder}), while the other
four relevant qubits (2-5) and the remaining unneeded physical qubits are
initially prepared in the computational basis state $|0\rangle$. The
approach for obtaining the operations $E_1$ and $E_2$ are shown in
Figs.~\ref{E1} and \ref{E2}, respectively. The distant GHZ operation depicted
in (b) between qubits 1-5 and qubit 6 is shown in Fig.~\ref{E3}.
When we perform all the operations in (a) and (b) including
several single-qubit operations shown in Fig.~\ref{Encoder}, the
final state is $|\psi'_{LP}\rangle_{AB}$ in Eq.~(\ref{LP01}).}
\end{figure}

\subsection{Two-qubit encoded cluster states}

The quantum circuit for the encoded teleportation is shown in
Fig.~\ref{Teleport}, but this cannot be directly implemented in systems
characterized by global entangling operations because it requires entanglement
gates between distant qubits. An appropriate strategy for
producing a logical-physical cluster state in a 2D square system is shown in
Fig.~\ref{E1E2GHZ}. All of the sites colored blue are qubits that serve as
ancillae, and are initialized in $|0\rangle$; the white dots are physical
qubits comprising logical qubits $A$ and $B$, labeled from 1 to 5
and 6 to 10 in Fig.~\ref{E1E2GHZ}, respectively. The encoding of
qubits in $A$ requires single-qubit operations on the physical
qubits, and two graph circuits: the GHZ$_5$ and the pentagon
operator $C_{1-5}^{\pentagon}$, both shown in Fig.~\ref{Encoder}.
As shown below, these can each be performed independently with
two global entangling gates (in the vertical and horizontal
directions) using the nearby ancillary qubits.
% , as shown in Appendix~\ref{E1} and \ref{E_2}.
The logical-physical cluster state is finally formed with the GHZ$_6$ operator
$\prod_{i=1}^5CZ_{i,6}$ shown in Fig.~\ref{E1E2GHZ}(b), and implemented
explicitly below using three global entangling operations
% in Appendix~\ref{GHZ3}
The logical-physical cluster state can therefore be obtained using a total of
seven global entangling operations.

\begin{figure}[t]
\hspace{-1.8cm}
\includegraphics[width= 6cm,angle=-90]{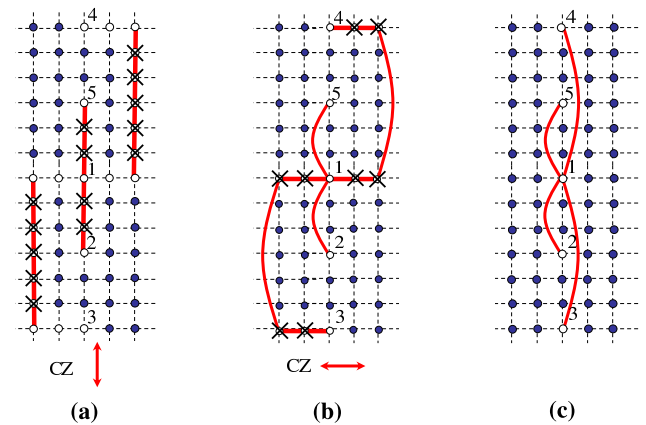}
\caption{ \label{E1} Two global entangling operations are required to build
the $E_1$ circuit. (a) Blue and white dots are initialized in states
$|0\rangle$ and
$|+\rangle$, respectively. After a vertical global $CZ$ operation (red lines),
various qubits are measured in the $X$ basis (denoted by the $\times$ symbol).
(b) After a horizontal global $CZ$ operation and further measurements, one
obtains the desired outcome (c).}
\end{figure}

The implementation of the four entangling operations in the $E_1$ circuit is
depicted in Fig.~\ref{E1}.
All of the qubits are initialized either in the states $|0\rangle$ (blue) or
$|+\rangle$ (white). A single global entangling operation is then performed
in the vertical direction, which results in $CZ$ gates between each pair of
white qubits (shown as straight red lines in the figure). Measurements in the
$X$ basis (depicted as crosses in the figure) are then performed on selected
ancillae in order to produce a $CZ$ gate between distant qubits (shown as
curved red lines in the figure). Because each measurement teleports an $X^mH$
gate to the neighboring qubit, where $m=\{0,1\}$ denotes the measurement
outcome, one must measure an even number of qubits between each pair.
Alternatively, one could choose to measure an odd number of intervening
qubits, but then a Hadamard gate would have to be applied manually to any
qubit at the end of the distant $CZ$ bond. This latter approach would require
fewer ancillae, but unnecessarily complicates the present discussion. A
subsequent entangling operation in the horizontal direction, followed again by
an even number of $X$ measurements and the application of appropriate
single-qubit gates, results in the four GHZ$_5$ entangling operations needed
for the $E_1$ circuit.

\begin{figure}[t]
\hspace{-1.8cm}
\includegraphics[width= 6cm,angle=-90]{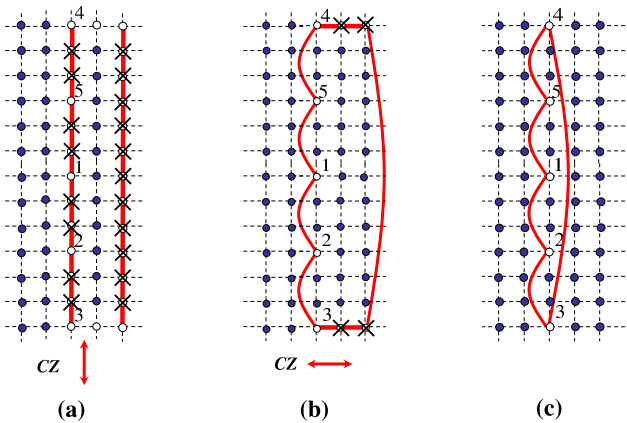}
\caption{ \label{E2} Two global entangling operations are also
required to build the $E_2$ circuit. (a) Blue and white dots are
initialized in states $|0\rangle$ and $|+\rangle$, respectively.
After a vertical global $CZ$ operation (red lines), various
qubits are measured in the $X$ basis (denoted by the $\times$
symbol). (b) After a horizontal global $CZ$ operation and further
measurements, one obtains the desired outcome (c).}
\end{figure}

The five entangling gates $C^{\pentagon}_{1-5}$ in the circuit $E_2$ can be
implemented in an analogous fashion, as shown in Fig.~\ref{E2}. The qubits
are initialized in either $|0\rangle$ or $|+\rangle$, a global vertical
entangling operation is performed, followed by various measurements and
single-qubit gates to compensate for byproduct operators associated with
measurement outcomes $m=1$. A subsequent horizontal global entanglement
operation and further measurements yields the desired gate set. Note that for
the last set of measurements, one cannot na\" \i vely apply single-qubit $X$
gates to compensate for byproduct operators, because these do not commute
with the $CZ$ operators; rather, the byproduct operators must be kept in mind
during the usual feed-forward process of the one-way computation.

The final task is to create the GHZ$_{1-5,6}$ operation between each of five
qubits of the logical qubit $A$ and physical qubit 6,
$\prod^{5}_{n=1} CZ_{n,6}$. While the approach is again similar to the methods
described above to generate entanglement between remote qubits in circuits
$E_1$ and $E_2$, in the present case one rather requires three global
entangling gates. This might at first seem surprising, because one requires a
set of entangling gates that is almost identical to those needed for $E_1$.
The reason is that qubit 6 needs to be entangled with five other qubits,
whereas in $E_1$ qubit needed to entangle with only four other qubits. With a
square lattice geometry where each qubit has four neighbors, entangling a
central qubit to four other qubits is straightforward, but not to five.

\begin{figure}[t]
\hspace{-1.7cm}
\includegraphics[width= 5.5cm,angle=-90]{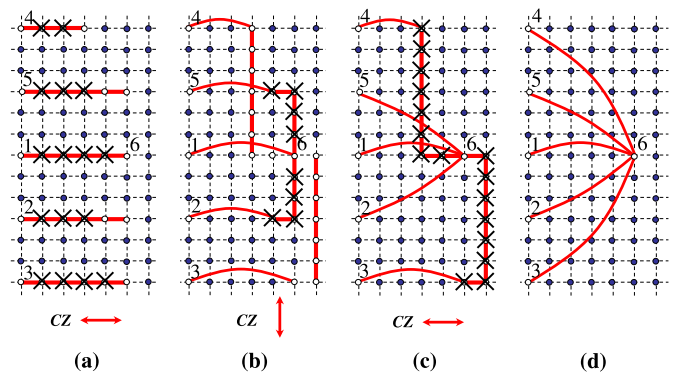}
\vspace{-0.5cm} \caption{ \label{E3} Three global entangling
operations are required to generate the GHZ$_6$ entangling gates
shown in Fig.~\ref{Teleport}. (a) Blue and white dots are
initialized in states $|0\rangle$ and $|+\rangle$, respectively.
After a horizontal global $CZ$ operation (red lines), various
qubits are measured in the $X$ basis (denoted by the $\times$
symbol). (b) After re-initialization of various qubits a vertical
global entangling gate is performed, and various measurements are
made. (c) A subsequent re-initialization, vertical entangling
operation, and set of measurements yields the desired outcome
(d).}
\end{figure}

The scheme for implementing the GHZ$_{1-5,6}$ operations in the
lattice is shown in Fig.~\ref{E3}. After suitable initialization of qubits, a
horizontal global entangling gate is
applied. After suitable measurements and applied single-qubit
unitaries, a vertical global entangling gate is applied. Again
making measurements and applying unitaries, one obtains
entanglement between qubit 6 and qubits 1, 2, and 5. In order to
generate entanglement between qubit 6 and qubits 3 and 4, one
requires one more vertical global entangling operation and
further measurements/unitaries.

With the approaches discussed above, one requires a total of seven global
entangling operations, in addition to single-qubit measurements and unitaries,
in order to produce the logical-physical cluster state $|\psi'_{LP}\rangle$.
The full quantum teleportation circuit shown in Fig.~\ref{Teleport} also
requires the decoding of logical qubit $A$ and the encoding of logical qubit
$B$, followed by error-syndrome and gate-teleportation measurements. Once the
information has been teleported, it will be encoded in logical qubit $B$,
because the encoding operation $E_2E_1$ will already have been carried out.
The information and a logical gate can then be teleported back to $A$ by
replacing the indices $1-5$ with $6-10$, etc. Thus, each subsequent
teleportation requires decoding $B$ ($A$), encoding $A$ ($B$), and performing
the GHZ$_{6-10,1}$ (GHZ$_{1-5,6}$) entangling gates.

One nice feature of this proposal is that the decoding of $A$ ($B$) can be
performed simultaneously with the encoding of $B$ ($A$). Both $E_1$ and $E_2$
are effected by a horizontal global entangling gate followed by a vertical one.
Furthermore, the qubits in $A$ and the ancillae used to mediate the distant
$CZ$ gates among them are well-separated from their counterparts in $B$. Thus,
each encoded gate teleportation requires a total of seven global entangling
operations. The only difference to Figs.~\ref{E1}-\ref{E3} is that physical
qubits comprising $A$ ($B$) are not re-initialized if the quantum information
is encoded in logical qubit $A$ ($B$).

\subsection{Four-qubit encoded cluster states}
\begin{figure}[t]
\hspace{-1.6cm}
\includegraphics[width= 5.8cm,angle=-90]{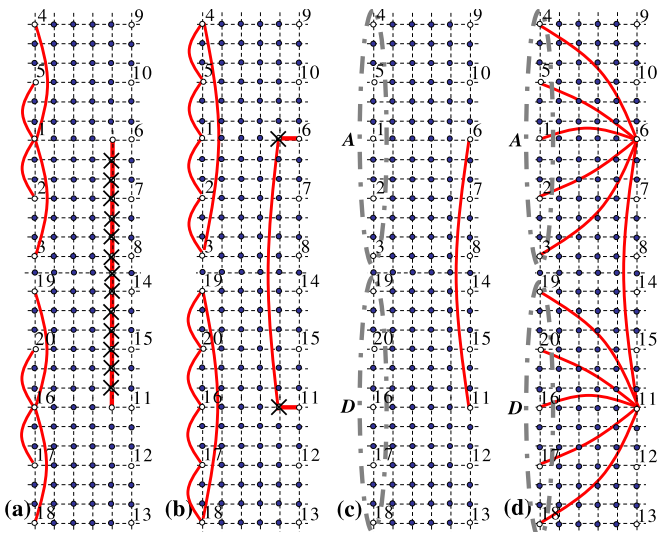}
\vspace{1.5cm} \caption{ Two logical qubits are denoted as $A$
(qubits 1 to 5) and $D$ (16 to 20). (a) During operation $E_1$ in
qubit $A$ and $D$, a vertical global CZ operation is
simultaneously created on the right side, and the middle qubits
are measured in the $X$ basis. (b) During operation $E_2$ in
qubit $A$ and $D$, a horizontal global CZ operation is performed,
and two auxiliary qubits are measured in $X$ basis. (c) While
logical qubits are successfully made in $A$ and $D$ (see the grey
dot-dash ovals), a distant CZ operation is created between qubits
6 and 11. (d) Performing GHZ$_6$ operation between logical qubit
$A$ and 6 ($D$ and 11), a logical-physical four-qubit cluster
state is created. \label{4LogicCS} }
\end{figure}

To build the horseshoe graphs $\sqsubset$ and $\sqsupset$ of the
computational cluster state, we need to perform only a distant
$CZ$ operation between the central qubit of a logical qubit and
that of the other logical qubit. As shown in Fig.~\ref{4LogicCS},
four logical qubits are denoted as $A$ (qubits 1 to 5), $B$
(qubits 6 to 10), $C$ (qubits 11 to 15), and $D$ (16 to 20). Their
central information qubit are called qubit 1, 6, 11, and 16, and
prepared in states $|\psi \rangle_{1} = \alpha |0 \rangle_{1} +
\beta |1 \rangle_{1}$, $|+\rangle_6$, $|+\rangle_{11}$, $|\phi
\rangle_{16} = c |0 \rangle_{16} + d |1 \rangle_{16}$,
respectively. In Fig.~\ref{4LogicCS}(a), a vertical global CZ
operation (a red straight line) is created, and the middle qubits
are measured in the $X$ basis (denoted by $\times$ symbol) during
operation $E_1$ on qubits $A$ (1 to 5) and $D$ (16 to 20).
Similarly, during operation $E_2$ on qubits $A$ and $D$, a
horizontal global CZ operation is made between qubit 6 (11) and
the auxiliary qubit and the two extra qubits are measured in $X$
basis as shown in Fig.~\ref{4LogicCS}(b). Then, in
Fig.~\ref{4LogicCS}(c), two logical qubits ($A$ and $D$) are
built (see the grey dot-dash lines) while a physical two-qubit
cluster state is made between qubits 6 and 11. This state yields
\begin{eqnarray}
{1\over\sqrt{2}} \left[ |\psi^L \rangle_A (|0 \rangle_6 |+
\rangle_{11}+ |1 \rangle_6 |- \rangle_{11}) |\phi^L
\rangle_D\right].
\end{eqnarray}
In Fig.~\ref{4LogicCS}(d), the state becomes a logical-physical
four-qubit cluster state with quantum information $|\psi^L
\rangle$ and $|\phi^L \rangle$ after the GHZ$_6$ operation between
logical qubit $A$ and physical qubit $6$ ($D$ and $11$). Finally,
when encoding $E_2 E_1$ is performed in qubits $B$ and $C$, the
final state yields $CZ^L_{AB}CZ^L_{BC}CZ^L_{CD}$ $|\psi^L
\rangle_A |+^L \rangle_B |+^L \rangle_C |\phi^L \rangle_D$ as the
logical horseshoe graph $\sqsupset$.

\subsection{Toward fault tolerant 1WQC}

The approach to one-way quantum computation described above has
several advantages over the standard cluster-state model. First,
the two-column approach means that physical qubits are prepared
quickly and then immediately used. In this way, the accumulation
of errors on resource qubits is minimized. Second, the quantum
information is protected during a portion of the protocol, using
the smallest possible QECC and the smallest number of entangling
operations. In this way, single qubit errors are detected and
corrected before they can be propagated by quantum teleportation.
Third, in many physical implementations most of the entanglement
gates required can be parallelized (for example, the encoding of
multiple logical qubits can be performed simultaneously with the
decoding of others), significantly lowering the number of
required operations.

These advantages notwithstanding, the procedure is only weakly tolerant of
single-qubit depolarizing errors. The information is only protected
during the small temporal window immediately following the gate teleportation.
%Major modifications would be required in order to ensure full fault-tolerance.
That said, there are several small changes to the above protocol that would
provide significant improvements to the protection of the quantum information,
and furthermore point the way toward developing a fault-tolerant 1WQC protocol
with suitably modified cluster states.

The first modification is to redesign the sequence of operations in order to
protect the quantum information from single-qubit errors most of the time. In
the current approach (see Fig.~\ref{Teleport}), the logical qubit A of a
logical-physical two-qubit cluster state, Eq.~(\ref{LP01}), is decoded at the
same time as logical qubit B is encoded. This has the advantage of minimizing
the total number of
entangling operations needed in systems where these can be performed in
parallel. The disadvantage is that the quantum information residing on qubit A
is susceptible to error during this process, as it is during the syndrome
measurements, possible attendant single-qubit unitaries, and the gate
teleportation measurement. It would be preferable if most if not all of these
operations could all be performed on encoded information; and furthermore that
they would be carried out fault-tolerantly.

To improve this scenario, qubit B could be encoded independently while the
quantum information continues to reside in A, which in systems characterized
by global entangling operations
requires a total of four global entangling gates, as shown in Figs.~\ref{E1}
and \ref{E2}. This approach has the definite advantage that the integrity of
the encoding can be verified before the quantum information in A is teleported
to B. As in the original protocol~\cite{DiVin96}, one needs to construct a
four-qubit GHZ state proximal to the logical state being tested; then four of
the physical qubits comprising the logical state are each entangled with one
qubit of the GHZ state, followed by syndrome measurements on the GHZ qubits.
The process is repeated four times to verify the logical state. Any error
discovered can be repaired by performing the appropriate single-qubit operation
on the logical state. Assuming perfect GHZ states (more on this below), the
procedure is fault-tolerant, because of the relationship between the QECC
stabilizer and the entangling operations between the logical and GHZ states.
The full procedure should be performed several times in order to have
confidence that the error syndrome is properly diagnosed.

Unfortunately, encoding qubit B before entangling with A modifies
Fig.~\ref{Teleport}, in that the unitary $E_2E_1$ is applied to qubits 6-10
{\it before} the GHZ$_{1-5,6}$ operation. It is not difficult to show that
this means that the full GHZ$_{1-5,6-10}$ set of twenty five entangling gates
depicted in Fig.~\ref{10qubit01} would instead need to be applied on the
physical qubits. This is the main apparent drawback of employing the
five-qubit QECC: because it is not a CSS code, a transversal $CZ$ operation
(i.e.\ a direct logical $CZ$ gate between encoded qubits) is not possible.
While the logical entangling gate cannot be applied fully fault-tolerantly
without a large overhead in terms of ancillary qubits and operations, certain
steps can nevertheless be verified.

The required set of gates can be simplified by noting that the graph state
$|g\rangle\equiv\prod_{i=1}^5\prod_{j=6}^{10}CZ_{i,j}|+\rangle^{\otimes 10}$
is unitary equivalent to the graph state
$|g'\rangle\equiv CZ_{1,6}\prod_{i=2}^5CZ_{1,i}\prod_{j=7}^{10}CZ_{6,j}
|+\rangle^{\otimes 10}$
by edge complementation~\cite{Danielsen2008} (in this case between qubits 1
and 6). That is, one rather needs to first generate GHZ states on qubits
1-5 and 6-10, and then entangle together only one qubit from each set, for
a total of only nine entangling gates.
The edge-complementation equivalence does not apply directly to the qubits A
and B, however, because one of these is an encoded state possessing the
relevant quantum information, and is not the simple product state
$|+\rangle^{\otimes 5}$ assumed above.

One would rather insert ancillary qubits (labeled $1'$ through $5'$) just to
the right of logical qubit A, oriented vertically and
aligned horizontally with qubits 1 through 5; a similar set would be inserted
to the left of logical qubit B (labeled $6'$ through $10'$). GHZ states
$|5GHZ\rangle_{1'-5'}$ and $|5GHZ\rangle_{6'-10'}$ are prepared
using the procedure shown in Fig.~\ref{E1}. The two GHZ states can then be
entangled with each other using a single horizontal entangling operation and
two additional ancillae, and the state depicted in Fig.~\ref{10qubit01} can
be obtained by local operations. Logical qubits A and B are then entangled by
applying a single horizontal global entangling gate, which entangles all qubit
pairs $i$ and $i'$, $1\leq i,i'\leq 5$ through intermediate ancillae, followed
by X measurements of qubits $i$.

The four and five-qubit GHZ states that are required to fault-tolerantly
verify the encoded single qubit state and to mediate the logical qubit-qubit
entangling operation, respectively, must themselves be verified prior to their
use. In practise, one simple requires one additional physical qubit, which is
repeatedly entangled with the GHZ state and then measured. If the measurement
outcome indicates an error, the procedure is repeated. The ancilla would be
a physical qubit directly above the GHZ qubits, so only one vertical entangling
operation is needed for each verification measurement. A similar approach was
recently proposed in ion traps~\cite{Oi2006}.

All logical states and ancillary GHZ states can be verified fault-tolerantly
after their (non fault-tolerant) construction. Thus, all operations in the
1WQC scheme discussed above can in principle be performed fault-tolerantly,
except for the single entangling operation that links the ancillary five-qubit
GHZ states to the logical qubits and to each other. This can be performed using
fault-tolerant constructions, but is not in itself fault-tolerant because the
fidelity of the result cannot be tested. In principle, this operation also
could be made fault-tolerant through the use of additional ancillae and
verification, but at the cost of completely repeating the construction of the
logical qubit not encoding the quantum
information. In practise, it is reasonable to assume that the time taken to
perform these entanglement operations, during which the quantum information
would be entangled fault-tolerantly but unverifiably, would be small compared
to the time needed to perform the other operations.

\section{Summary and remarks}
\label{conclusions}

An approach to 1WQC has been presented that explicitly incorporates quantum
error correction as a way to minimize the propagation of errors during the
computational process. This proposal has two key features. First, the cluster
states are only constructed two columns at a time, so that physical qubits will
not have a sufficient amount of time to undergo significant decoherence prior
to measurement. Second, QECC states are prepared in each column, where a
logical qubit is equivalent to a five-qubit graph state, and a logical
two-qubit cluster state is represented by a graph state of ten qubits. In this
way, errors can be detected and corrected prior to the main gate teleportation.
Each teleportation requires several procedures on the two logical qubits,
including encoding, decoding, GHZ$_6$ quantum circuits. Prior to the encoded
teleportation, the error syndrome is checked by single-qubit measurements on
four of the five physical qubits comprising the first logical qubit, and the
corrected information is teleported to the next logical qubit.

Following the description of the fundamental protocol, a physical
implementation is discussed for systems characterized by global entangling
operations, such as ultracold atomic gases in 2D optical lattices. A procedure
for constructing logical cluster states (logical two-qubit and four-qubit
states) using several global $CZ$ operations and single-qubit measurements is
shown explicitly, with an eye on minimizing the number of physical (ancillary)
qubits and total operations.

While the main scheme for error-correcting 1WQC discussed in the
manuscript allows for error detection and correction, it is not
fault-tolerant. An improved scheme is outlined that begins to
address this issue, at the cost of additional ancillae and
operations. This improved scheme incorporates fault-tolerant
elements, but only hints at a fully fault-tolerant approach to
one-way quantum computation based on quantum error correction.
For example, the logical states are constructed using unprotected
single-qubit and entangling gates, logical qubits are entangled
by non-transversal and unverifiable operations, and concatenating
the logical qubits in systems characterized by global entangling
operations seems daunting. 

In the circuit model, a
concatenation method for five-qubit QECC has been already
mathematically studied~\cite{GottesmanPhD}. In our scheme, the
first level of concatenation costs 25 physical qubits (five
pentagon qubits) and 50 physical $CZ$ operations among them using
edge-complementation equivalence. Because of the mathematical complexity of
the logical $CZ$ operation in the first level concatenation, let alone
determining how one would go about implementing it with physical qubits in
periodic lattices, a full investigation into how to implement concatenation
with this approach is beyond the scope of the present work.
Another fruitful extension of our work, concatenating two different QECCs and
using larger codes to correct more noisy models, will also be needed in order
to develop a more complete theory of practical fault-tolerant 1WQC.
These issues, and a full analysis of the associated error thresholds, will be
addressed in future work.

\begin{acknowledgements}
The authors acknowledge fruitful conversations with Travil Beals.
J.J.~acknowledges iCORE and the Natural Sciences and Engineering
Research Council of Canada. This work was supported by the
Natural Sciences and Engineering Research Council of Canada, the
Alberta Informatics Circle of Research Excellence, and the
Mathematics of Information Technology and Complex Systems'
Quantum Information Processing project.

\end{acknowledgements}

\end{document}